# Experimental demonstration of steady-state dynamics of three-level quantum heat engine using superconducting quantum circuits

Gao-xiang Deng[a], Haoqiang Ai[b], Wei Shao[b], Yu Liu[a], Zheng Cui[b,*]

a Institute of Thermal Science and Technology, Shandong University, Jinan, 250061, P.R. China

b Shandong Institute of Advanced Technology, Jinan, 250100, P.R. China

*Correspondence: zhengc@sdu.edu.cn

## Abstract

The three-level system represents the smallest quantum system capable of autonomous cycling in quantum heat engines. This study proposes a method to simulate the steady-state dynamics of a three-level quantum heat engine by designing and implementing superconducting quantum circuits. Following error mitigation, the outcomes from the quantum circuit model designed in this study, when executed on a real quantum device, closely align with theoretical predictions, thereby validating the effectiveness of the circuit model. This study offers a novel approach for investigating three-level quantum heat engines, enabling the verification of theoretical research findings while also reducing the complexity and cost of experimental procedures.

# 1 Introduction

A heat engine is traditionally understood as a macroscopic device that transforms thermal energy into mechanical work. However, the ongoing trend of device miniaturization has begun to expose quantum effects. Quantum heat engines (QHEs), which operate at the quantum scale to convert energy, have emerged in response to this trend. The concept of QHEs was first introduced in the 1960s [1, 2], and its theoretical underpinning, quantum thermodynamics [3], has been progressively refined following the maturation of open quantum systems theory [4]. Bolstered by the advancements in quantum thermodynamics and the potential applications revealed by device miniaturization, research on QHEs has seen a significant surge over the past two decades [5].

An open question in the field of QHEs is whether quantum effects can enhance performance, a phenomenon often referred to as "quantum advantage". Theoretically, some researchers argue that "quantum friction" [6-10] could potentially decrease the performance of QHEs. Conversely, others propose that quantum effects such as quantum coherence [11-17], quantum entanglement [18-20], energy level degeneracy [17, 21, 22], many-body cooperation [23-26], and squeezed thermal reservoirs [27-29] could enhance the performance of QHEs with appropriate control.

Advancements in experimental technology have enabled researchers to maintain and observe quantum effects and construct QHEs on various systems including photons [30, 31], negatively charged nitrogen vacancy (NV$^-$) centers in diamond [32-34], particle pair spins [35-38], electrons [39-42], and cold atoms [43, 44] or ions [45-48]. Some of these experimental studies have suggested that quantum effects can indeed enhance the performance of QHEs [33, 34, 36, 41, 45]. However, it remains unclear whether this performance enhancement necessitates the consumption of additional energy, such as the use of non-thermal resources to squeeze thermal reservoirs for performance enhancement [49, 50].

One approach to eliminate the influence of additional energy is to apply "autonomous operation" [51-53] to the cycle of QHEs. Recent research [54] has indicated that a three-level system, as opposed to a qubit, represents the smallest quantum system capable of achieving autonomous thermodynamic cycling. This can be accomplished solely by relying on the temperature difference between the cold and hot reservoirs, without the need for external control.

Theoretical researches on three-level QHEs also suggest that quantum effects could potentially enhance the performance of QHEs [12, 17, 55-62]. Correspondingly, experimental studies have reported enhanced performance of three-level QHEs [33], provided the thermal stroke time is less than the dephasing time. These findings indicate that three-level QHEs hold significant potential for achieving quantum advantage. However, the implementation of theoretical results and the high cost of experiments present considerable challenges.

Cloud quantum devices, such as those provided by the IBM Quantum Platform, offer a solution to these challenges by allowing researchers to submit theoretical models for execution, thereby significantly reducing experimental costs. Furthermore, IBM's cloud quantum devices have been shown to exhibit robust performance when executing open quantum system models [63, 64]. Consequently, open quantum system models, including quantum synchronization [64-66], quantum steering [67], two-level QHE [68], and the dissipative two-site Hubbard model [69], have been successfully implemented on the IBM Quantum Platform. Despite these advancements, three-level QHE models remain relatively unexplored.

This manuscript presents a scheme to run a three-level QHE on superconducting quantum circuits. Firstly, we introduce the physical model of the three-level QHE and accordingly construct its steady-state quantum circuit model with suitable simplification. This circuit model is then executed on real quantum devices via the IBM Quantum Platform. Based on the execution results, we perform error mitigation and validate the circuit model by comparing the calibrated results with theoretical predictions and simulation results. Subsequently, the impact of different initial states and the simplification are discussed. Finally, we summarize our findings.

## 2 Models and Methods

## 2.1 Circuit model

Fig. 1(a) gives the thermodynamic model of the three-level QHE, where the energy level transitions dictate the corresponding system dynamics. When the quantum system coupled with a hot reservoir at temperature $T_h$, a cold reservoir at temperature $T_c$, or an external field $V$ will lead a transition between $|0\rangle$ and $|2\rangle$, $|0\rangle$ and $|1\rangle$, or $|1\rangle$ and $|2\rangle$, respectively [70]. The dynamics of the three level QHE is governed by the Gorini-Kossakowski-Lindblad-Sudarshan (GKLS) equation [4, 71],

$$\partial_t \rho_S = -i[H_S, \rho_S] + \sum_{\alpha=c,h} \mathcal{D}_\alpha(\rho_S) \tag{1}$$

where $H_S$ is the Hamiltonian of the quantum system S, $\mathcal{D}_\alpha$ is the dissipator, subscript $c$ and $h$ represent the hot reservoir and cold reservoir, respectively.

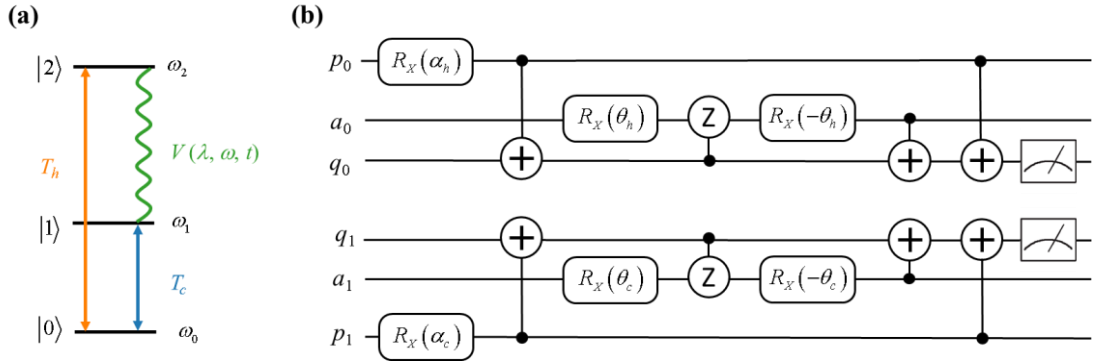

Fig. 1. (a) Energy levels of three-level QHE. $|0\rangle$, $|1\rangle$, and $|2\rangle$ are the eigenstates of the free Hamiltonian, and $\omega_0$, $\omega_1$, and $\omega_2$ are the corresponding eigenvalues. $T_h$ and $T_c$ are the temperature of hot and cold reservoir. The external field V are represented by the intensity $\lambda$, oscillating frequency $\omega$, and evolution time t. (b) Quantum circuit used to simulate the thermodynamics of the three-level QHE. The system states are represented by system qubit, $q_0$ and $q_1$. Correspondingly, the ancillary qubit and probability qubit for system qubits are denoted as $a_0$, $p_0$ and $a_1$, $p_1$, respectively. The quantum gates, which are applied to each qubit, facilitate the thermodynamics simulation of three-level. A detailed discussion on the quantum circuit will be provided in the main text.

Typically,

$$H_S = \begin{pmatrix} \omega_0 & 0 & 0 \\ 0 & \omega_1 & \lambda e^{i\omega t} \\ 0 & \lambda e^{-i\omega t} & \omega_2 \end{pmatrix} \tag{2}$$

for this three-level QHE. However, in order to facilitate the construction of quantum circuits, the phase of the non-diagonal elements in $H_S$ is assumed to be time-invariant, i.e.,

$$H_S = \begin{pmatrix} \omega_0 & 0 & 0 \\ 0 & \omega_1 & \lambda \\ 0 & \lambda & \omega_2 \end{pmatrix} \tag{3}$$

Concurrently, the basis of the density matrix is projected from original eigenstates ($|0\rangle$, $|1\rangle$, $|2\rangle$) to the diagonal eigenstates of $H_S$ ($|\varepsilon_0\rangle$, $|\varepsilon_1\rangle$, $|\varepsilon_2\rangle$) [72]. Meanwhile, it is postulated that there are no off-diagonal elements present in the initial density matrix [68]. Therefore, Eq. (1) can be wrote as:

$$\partial_t \rho_S = \sum_\alpha \mathcal{D}_\alpha(\rho_S) = \sum_{\alpha,\varepsilon} \mathcal{D}_\alpha^\varepsilon(\rho_S) \tag{4}$$

For the "resonant coupling" [73], the dissipation operator expressed as:

$$\begin{aligned}
\mathcal{D}_h^{\varepsilon_{20}}(\rho_S) &= \gamma_h(\varepsilon_{20})\cos^2\frac{\theta}{2}\begin{bmatrix} \rho_{2,2} & 0 & 0 \\ 0 & 0 & 0 \\ 0 & 0 & -\rho_{2,2} \end{bmatrix} \\
\mathcal{D}_h^{-\varepsilon_{20}}(\rho_S) &= \gamma_h(-\varepsilon_{20})\cos^2\frac{\theta}{2}\begin{bmatrix} -\rho_{0,0} & 0 & 0 \\ 0 & 0 & 0 \\ 0 & 0 & \rho_{0,0} \end{bmatrix} \\
\mathcal{D}_c^{\varepsilon_{10}}(\rho_S) &= \gamma_c(\varepsilon_{10})\cos^2\frac{\theta}{2}\begin{bmatrix} \rho_{1,1} & 0 & 0 \\ 0 & -\rho_{1,1} & 0 \\ 0 & 0 & 0 \end{bmatrix} \\
\mathcal{D}_c^{-\varepsilon_{10}}(\rho_S) &= \gamma_c(-\varepsilon_{20})\cos^2\frac{\theta}{2}\begin{bmatrix} -\rho_{0,0} & 0 & 0 \\ 0 & \rho_{0,0} & 0 \\ 0 & 0 & 0 \end{bmatrix}
\end{aligned} \tag{5}$$

where, $\tan\theta = 2\lambda/(\omega_2 - \omega_1)$, $\rho_{0,0} = \langle\varepsilon_0|\rho_S|\varepsilon_0\rangle$, $\rho_{1,1} = \langle\varepsilon_1|\rho_S|\varepsilon_1\rangle$, $\rho_{2,2} = \langle\varepsilon_2|\rho_S|\varepsilon_2\rangle$.

Subsequently, predicated on the "Generalized Amplitude Damping (GAD)" [68], the solution to Eq. (4) can be expressed as:

$$\rho_S(t+\Delta t) = \mathcal{E}_{\text{GAD-3}}[\rho_S(t)] \tag{6}$$

where, $\mathcal{E}_{\text{GAD-3}}$ is the GAD for the three-level QHE, and it's detailed expression is :

$$\mathcal{E}_{\text{GAD-3}} = p_h^{\downarrow}\mathcal{E}_h^{\downarrow} + p_h^{\uparrow}\mathcal{E}_h^{\uparrow} + p_c^{\downarrow}\mathcal{E}_c^{\downarrow} + p_c^{\uparrow}\mathcal{E}_c^{\uparrow} \tag{7}$$

In this context, the hot amplitude damping $\mathcal{E}_h^{\downarrow}$, hot amplitude pumping $\mathcal{E}_h^{\uparrow}$, cold amplitude damping $\mathcal{E}_c^{\downarrow}$, and cold amplitude pumping $\mathcal{E}_c^{\uparrow}$ are given by:

$$\begin{aligned}
\mathcal{E}_h^{\downarrow}(\rho_S) &= M_h^0 \rho_S (M_h^0)^{\dagger} + M_h^1 \rho_S (M_h^1)^{\dagger} \\
\mathcal{E}_h^{\uparrow}(\rho_S) &= M_h^2 \rho_S (M_h^2)^{\dagger} + M_h^3 \rho_S (M_h^3)^{\dagger} \\
\mathcal{E}_c^{\downarrow}(\rho_S) &= M_c^0 \rho_S (M_c^0)^{\dagger} + M_c^1 \rho_S (M_c^1)^{\dagger} \\
\mathcal{E}_c^{\uparrow}(\rho_S) &= M_c^2 \rho_S (M_c^2)^{\dagger} + M_c^3 \rho_S (M_c^3)^{\dagger}
\end{aligned} \tag{8}$$

Their corresponding probability are:

$$\begin{aligned}
p_h^{\downarrow} &= \frac{g_h^{\downarrow}}{g_h^{\downarrow} + g_h^{\uparrow}} \\
p_h^{\uparrow} &= \frac{g_h^{\uparrow}}{g_h^{\downarrow} + g_h^{\uparrow}} \\
p_c^{\downarrow} &= \frac{g_c^{\downarrow}}{g_c^{\downarrow} + g_c^{\uparrow}} \\
p_c^{\uparrow} &= \frac{g_c^{\uparrow}}{g_c^{\downarrow} + g_c^{\uparrow}}
\end{aligned} \tag{9}$$

The operators for hot reservoir are:

$$M_h^0 = \begin{bmatrix} 1 & 0 & 0 \\ 0 & 0 & 0 \\ 0 & 0 & \cos\theta_h \end{bmatrix}, M_h^1 = \begin{bmatrix} 0 & 0 & \sin\theta_h \\ 0 & 0 & 0 \\ 0 & 0 & 0 \end{bmatrix}$$
$$M_h^2 = \begin{bmatrix} \cos\theta_h & 0 & 0 \\ 0 & 0 & 0 \\ 0 & 0 & 1 \end{bmatrix}, M_h^3 = \begin{bmatrix} 0 & 0 & 0 \\ 0 & 0 & 0 \\ \sin\theta_h & 0 & 0 \end{bmatrix} \tag{10}$$

And for cold reservoir, they are:

$$M_c^0 = \begin{bmatrix} 1 & 0 & 0 \\ 0 & \cos\theta_c & 0 \\ 0 & 0 & 0 \end{bmatrix}, M_c^1 = \begin{bmatrix} 0 & \sin\theta_c & 0 \\ 0 & 0 & 0 \\ 0 & 0 & 0 \end{bmatrix}$$
$$M_c^2 = \begin{bmatrix} \cos\theta_c & 0 & 0 \\ 0 & 1 & 0 \\ 0 & 0 & 0 \end{bmatrix}, M_c^3 = \begin{bmatrix} 0 & 0 & 0 \\ \sin\theta_c & 0 & 0 \\ 0 & 0 & 0 \end{bmatrix} \tag{11}$$

Since qubit is a two-level system, in order to simulate the three-level QHE, the Hilbert space of the QHE needs to be mapped onto the logical Hilbert space of the quantum circuit's system qubits as follows [65]:

$$|\varepsilon_0\rangle=|00\rangle, |\varepsilon_1\rangle=|01\rangle, |\varepsilon_2\rangle=|10\rangle, |X\rangle=|11\rangle \qquad (12)$$

At this juncture, the evolution of the three-level QHE can be approximated using quantum circuit depicted in Fig. 1(b). This circuit is composed of six qubits. Specifically, $q_0$ and $q_1$ are system qubits, which are utilized to map the dynamics of the three-level QHE. Ancillary qubits, $a_0$ and $a_1$, are employed to implement amplitude damping on system qubits, i.e., implement $\mathcal{E}_h^{\downarrow}$ on $q_0$ and $\mathcal{E}_c^{\downarrow}$ on $q_1$, respectively. Probability qubits, $p_0$ and $p_1$, govern the of damping (pumping) probabilities of $q_0$ and $q_1$, respectively. For instance, $a_0$ implements damping operation ($\mathcal{E}_h^{\downarrow}$) on $q_0$ when the state of $a_0$ is $|0\rangle$, or pumping operation ($\mathcal{E}_h^{\uparrow}$) when the state of $a_0$ is $|1\rangle$. All the gate parameters in this circuit are given in Table 1.

Table 1. Gate parameters in the quantum circuits used for the simulation of the three-level QHE. Here, $\Delta t$ is the time step for simulation.

| Parameters | value |
|---|---|
| $\alpha_h$ | $2\cos^{-1}\left(\sqrt{p_h^\downarrow}\right)$ |
| $\alpha_c$ | $2\cos^{-1}\left(\sqrt{p_c^\downarrow}\right)$ |
| $\theta_h$ | $\cos^{-1}\left[\exp\left(-\frac{\Delta t}{2}\left(g_h^\downarrow + g_h^\uparrow\right)\right)\right]$ |
| $\theta_c$ | $\cos^{-1}\left[\exp\left(-\frac{\Delta t}{2}\left(g_c^\downarrow + g_c^\uparrow\right)\right)\right]$ |
| $g_h^\downarrow$ | $\gamma_h(\varepsilon_{20})\cos^2\frac{\theta}{2}$ |
| $g_h^\uparrow$ | $\gamma_h(-\varepsilon_{20})\cos^2\frac{\theta}{2}$ |
| $g_c^\downarrow$ | $\gamma_c(\varepsilon_{10})\cos^2\frac{\theta}{2}$ |
| $g_c^\uparrow$ | $\gamma_c(-\varepsilon_{10})\cos^2\frac{\theta}{2}$ |
| $\gamma_h(-\varepsilon_{20})$ | $\gamma_h(\varepsilon_{20})e^{-\beta_h\varepsilon_{20}}$ |
| $\gamma_c(-\varepsilon_{10})$ | $\gamma_c(\varepsilon_{10})e^{-\beta_c\varepsilon_{10}}$ |
| $\varepsilon_{20}$ | $\frac{\omega_2+\omega_1}{2}+\sqrt{\left(\frac{\omega_2-\omega_1}{2}\right)^2+\lambda^2}$ |
| $\varepsilon_{10}$ | $\frac{\omega_2+\omega_1}{2}-\sqrt{\left(\frac{\omega_2-\omega_1}{2}\right)^2+\lambda^2}$ |

## 2.2 Error mitigation

The raw data obtained from running specified quantum circuits on real quantum devices often deviates significantly from theoretical calculations due to environmental noise, decoherence, quantum gate defects, measurement errors, and circuit topology. Generally, theoretical models cannot account for all environmental noise and accurate decoherence, leading to deviations. Simultaneously, imperfect quantum gates and the precision of measurements on real quantum devices also affect the raw data. Furthermore, it is necessary to compile and topologically optimize the designed quantum circuit prior to execution on real quantum devices, in order to minimize both runtime and gate errors.

Here, we employ the method of general error mitigation (GEM) [74] to the system qubits to correct the raw data. The main steps of GEM are as follows:

(1) Run the original circuit: Obtain the original probability of the joined states of system qubits, i.e., raw data, $\vec{v}$.

(2) Construct the calibration circuit: Divide the original circuit $C_{\text{org}}$ into two halves on average. The first half plus the inverse of the first half forms the first calibration circuit $C_{\text{clb}}^1$. Similarly, the second half plus the inverse of the second half forms the second calibration circuit $C_{\text{clb}}^2$.

(3) Run the calibration circuit: Prepare all possible $2^N$ initial states, twice. Measure the calibration circuit $C_{\text{clb}}^1$ and $C_{\text{clb}}^1$ with $2^N$ different initial states to obtain the calibration matrices $M_{\text{clb}}^1$ and $M_{\text{clb}}^1$, respectively. The average of these matrices gives the calibration matrix $M_Q = \left(M_{\text{clb}}^1 + M_{\text{clb}}^2\right)/2$.

(4) Calibrate the error: Apply least squares to obtain the calibrated data, i.e., finding the calibrated data which minimized the following function:

$$f(x) = \sum_{i=1}^{2^N} \left(v_i - \left(M_Q \cdot \vec{x}\right)_i\right)^2 \tag{13}$$

Here, $N$ is the number of system qubits, $\vec{x}$ is the calibrated data, $M_Q$ is the calibration matrix, and,

$$\begin{aligned}\vec{v} &= \left(v_1,\ v_2,\ \cdots,\ v_{2^N}\right)^{\text{T}} \\ \vec{x} &= \left(x_1,\ x_2,\ \cdots,\ x_{2^N}\right)^{\text{T}}\end{aligned} \tag{14}$$

For the purpose of error mitigation, the circuit in Fig. 1(b) is modified as shown in Fig. 2(a) and the corresponding calibration circuit is provided in Fig. 2(b). To avoid errors caused by the SWAP gate [69], we choose to reset the auxiliary qubits after the dashed line instead of introducing new qubits. Additionally, since the number of system qubits is 2, we need to prepare all 4 possible initial states for each system qubit of the calibration circuit. Following step (3) mentioned above, we can obtain the calibration matrix $M_Q$.

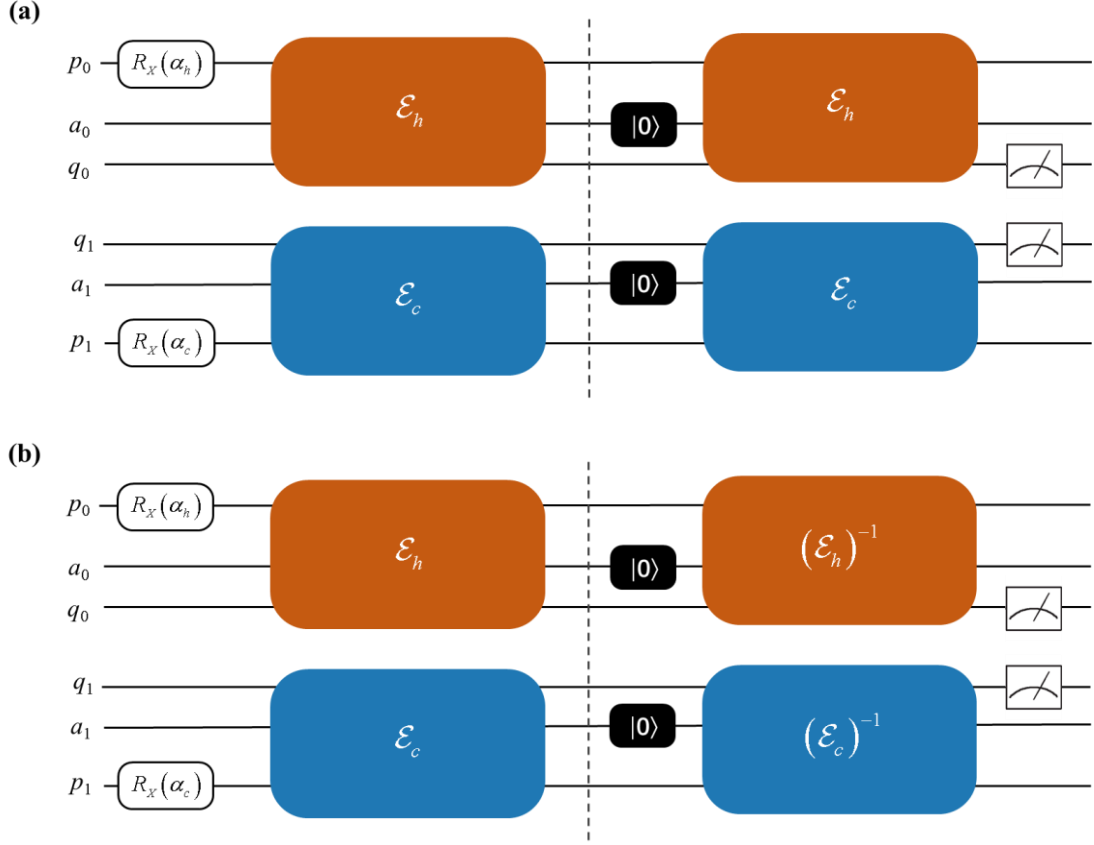

Fig. 2. (a) Schematic diagram of the simulation circuit modified for GEM. In this circuit diagram, the orange gate, $\mathcal{E}_h$, represents the dissipation of the system qubit $q_0$ under the influence of the hot reservoir, and the blue gate, $\mathcal{E}_c$, signifies the dissipation of the system qubit $q_1$ induced by the cold reservoir. The black gates following the dashed line indicates the reset gates to reset the states of the auxiliary qubits $a_0$ and $a_1$. (b) Calibration circuit. It's worth emphasizing that due to the symmetry of the modified circuit, specifically Fig. 2(a), its two calibration circuits, $C_{\text{clb}}^1$ and $C_{\text{clb}}^1$, are identical. Consequently, only one calibration circuit is presented here for simplicity and clarity. $(\mathcal{E}_h)^{-1}$ and $(\mathcal{E}_c)^{-1}$ are the inversed operation for $\mathcal{E}_h$ and $\mathcal{E}_c$, respectively. The detailed structure for $\mathcal{E}_h$, $\mathcal{E}_c$, $(\mathcal{E}_h)^{-1}$, and $(\mathcal{E}_c)^{-1}$ are given in Fig. 3.

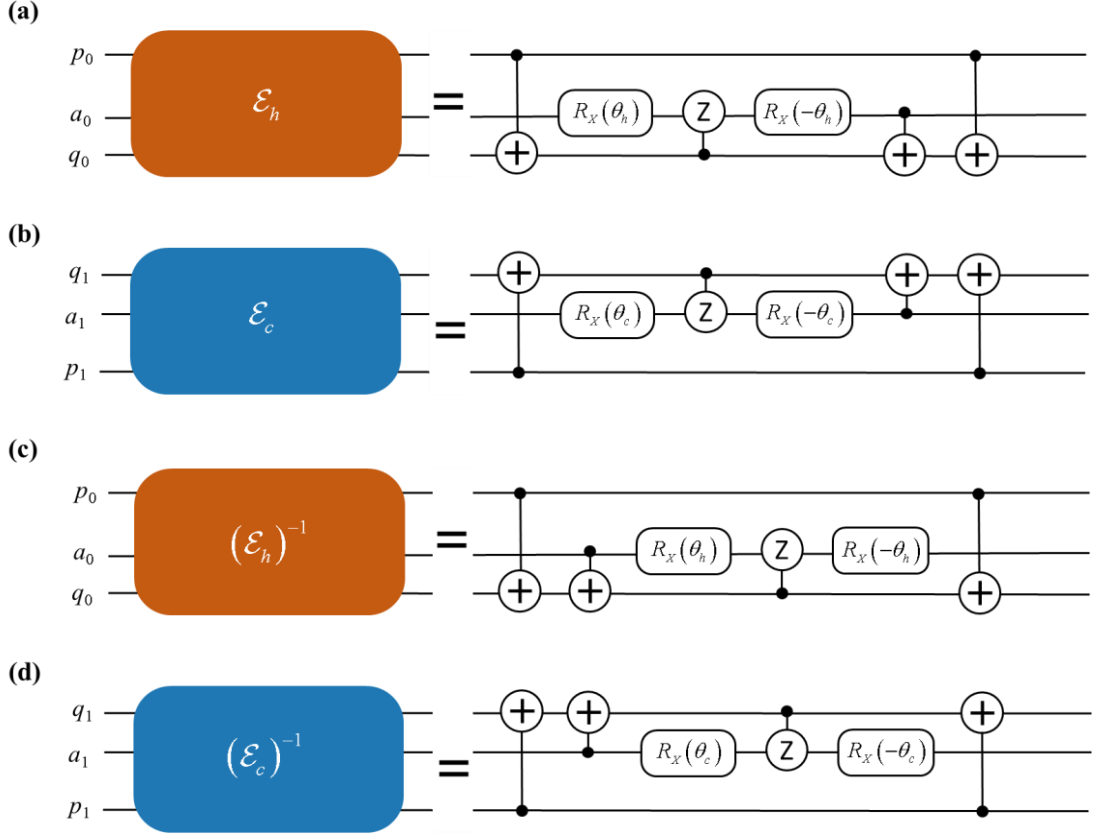

Fig. 3 The detailed structure for $\mathcal{E}_h$ (a), $\mathcal{E}_c$ (b), $(\mathcal{E}_h)^{-1}$ (c), and $(\mathcal{E}_c)^{-1}$ (d).

## 2.3 Methods

We employed *Qiskit* [75], a *Python* computing framework provided by IBM, for the construction of quantum circuits, circuits simulation, and circuits submission. Specifically, we selected the qasm_simulator as the simulation backend. As for the real backend, which refers to the real quantum device on the IBM Quantum cloud that executes the submitted circuits, we opted for ibm_lagos. Furthermore, we chose the *transpile* tool offered by *Qsikti*, setting the optimization_level to 2, to optimize the circuit topology. For the minimization problem in GEM, we employed the *minimize* package of the *Python* scientific computing tool, *scipy* [76].

## 3 Results and discussions

The physical parameters of the three-level QHE are detailed in Table 2. Each circuit, either conducted on ibm_lagos or simulated on qasm_simulator, comprises five runs, with each run consisting of 8192 shots.

**3.1 Experimental results**

The time evolution of populations for theoretical calculation and *Qiskit* simulation (via qasm_simulator) are given in Fig. 4(a). The theoretical predictions and simulation results exhibit a strong agreement. Meanwhile, Fig. 4(b) and 3(c) respectively present raw experimental data and calibrated data by GEM, using theoretical predictions (lines) as reference. As can be observed from the above figures, the calibrated data, after GEM, aligns more closely with the theoretical predictions compared to the raw data. However, as time progresses, $\rho_{0,0}$ for both raw data and calibrated data increases due to the relaxation of the system qubits. Overall，this concurrence validates the feasibility of the quantum circuit model depicted in Fig. 2(a) for simulating the dynamics of the three-level QHE.

Table 2. Physical parameters of the three-level QHE [73]. $\beta_h = 1/k_B T_h$ and $\beta_c = 1/k_B T_c$ are the inverse temperature of hot and cold reservoir, respectively. $\omega_{10} = \omega_1 - \omega_0$ and $\omega_{20} = \omega_2 - \omega_0$ are the difference of the original eigenfrequencies. For aspect calculations, setting $k_B = 1$.

| System parameters | value |
| --- | --- |
| $\beta_h \omega_{10}$ | 1 |
| $\beta_c \omega_{10}$ | 5 |
| $\omega_{20}/\omega_{10}$ | 2.5 |
| $\lambda/\omega_{10}$ | 0.5 |
| $\gamma_h(\varepsilon_{20})/\omega_{10}$ | 2 |
| $\gamma_c(\varepsilon_{10})/\omega_{10}$ | 2 |
| $\gamma_h(\varepsilon_{10})/\omega_{10}$ | 0 |
| $\gamma_c(\varepsilon_{20})/\omega_{10}$ | 0 |

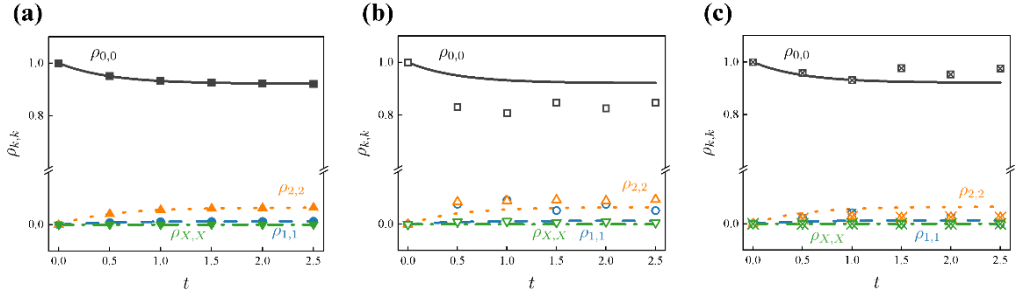

Fig. 4. Time evolution of the populations. (a)Theoretical predictions and simulation results. Black solid line (square), blue dashed line (circle), orange dotted line (triangle) and green dash-dotted line (inverted triangle) correspond to theoretical (simulation) populations of $\rho_{0,0}$, $\rho_{1,1}$, $\rho_{2,2}$, and $\rho_{X,X}$, respectively. (b) Theoretical predictions (lines) and raw experimental results (hollow symbols) from ibm_lagos. (c) Theoretical predictions (lines) and the GEM-calibrated data (hollow symbols with 'X') with raw data showing in panel (b). In the theoretical calculation, the range for $t$ spans from 0 to 2.5. The simulations and experiments are both conducted in 2 steps, with a time step from 0 to 1.25 in increments of 0.25.

**3.2 Dissipative stabilization**

Fig. 5 presents both the theoretical predictions and simulation results for the evolution of populations in various QHE initial states under dissipation. The simulation results across all three figures align with the theoretical predictions, thereby validating the efficacy of the quantum circuit model proposed in this study. Moreover, the final populations across different initial conditions are closely matched, confirming that the stability of the dissipative process is independent of the initial state.

A comparison of the illustrations in Fig. 5(a), 4(b), and 4(c) reveals that $\rho_{1,1}$ is larger in 4(b) and $\rho_{2,2}$ is larger in 4(c). This discrepancy is attributed to the limited evolution time (e.g., $t = 2.5$), which prevents the initial states from fully converging to the same final state. This inconsistency is further reflected in the differing slopes of the line segments in the illustrations of Fig. 5(b) and 4(c), where the blue line segment in 4(b) and the orange line segment in 4(c) indicate that $\rho_{1,1}$ and $\rho_{2,2}$ are still undergoing relatively rapid changes.

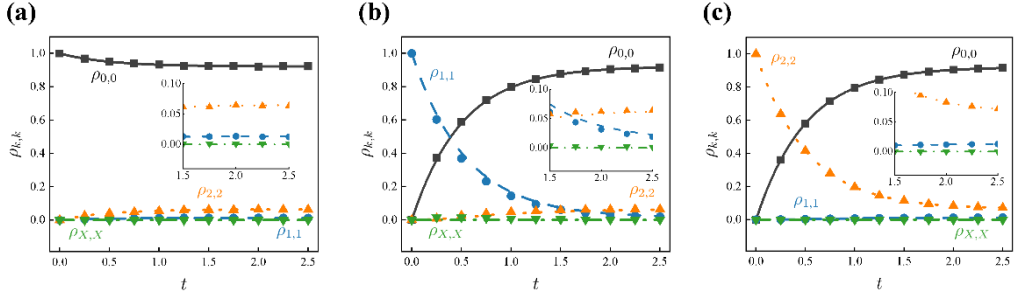

Fig. 5. The populations evolution of the three-level QHE in different initial states: $\rho_S(0)=|\varepsilon_0\rangle\langle\varepsilon_0|$ (a), $\rho_S(0)=|\varepsilon_1\rangle\langle\varepsilon_1|$ (b), and $\rho_S(0)=|\varepsilon_2\rangle\langle\varepsilon_2|$ (c). The line segments and points in these figures have the same interpretations as in Fig. 4. However, the time step increment for the quantum circuit simulation has been reduced from 0.25 to 0.125. The inset zoom provides a detailed view of the evolution of $\rho_{1,1}$, $\rho_{2,2}$, and $\rho_{X,X}$, within the time interval $t\in[1.5,2.5]$.

**3.3 Hamiltonian simplification**

In Section 2.1, we employed Eq. (3) to simplify the Hamiltonian of the three-level QHE, thereby facilitating the construction of a quantum circuit model. This section will explore the implications of this simplification on the evolution of populations.

Fig. 6 illustrates the theoretical calculation results for both the simplified and unsimplified models in different initial states. Firstly, the overlapping black line segments in the figures indicate that neither the simplification nor the initial state influences the evolution of $\rho_{0,0}$. This is because the simplification of Eq. (3) is essentially equivalent to setting $\omega=0$, which will affect the evolution of $\rho_{1,1}$ and $\rho_{2,2}$ but not $\rho_{0,0}$. For a detailed description of the populations evolution, please refer to Appendix B of Ref. [72].

However, both the simplification and the difference in the initial state significantly impact $\rho_{1,1}$ and $\rho_{2,2}$. In Fig. 6(a), the difference between $\rho_{1,1}$ and $\rho_{2,2}$ before and after simplification is small (approximately 0.02). In contrast, in Fig. 6(b) and 5(c), when $t\leq 1.5$, the difference is large (about 0.2), but when $t=2.5$, the difference becomes smaller (about 0.02). This is because when $\rho_{1,1}$ is too large, the simplification of Eq. (3) slows down the transition from $|\varepsilon_1\rangle$ to $|\varepsilon_2\rangle$, resulting in an overly large $\rho_{1,1}$ and an overly small $\rho_{2,2}$, as seen in Fig. 6(b). Similarly, when $\rho_{2,2}$ is too large, the simplification slows down the fallback process from $|\varepsilon_2\rangle$ to $|\varepsilon_1\rangle$, leading to the result in Fig. 6(c). Eventually, as the populations approach the steady state, the values of $\rho_{1,1}$ and $\rho_{2,2}$ become small and tend towards stability, reducing the impact of simplification on the results. This is evident in the evolution process of Fig. 6(a) and the final states of 5(b) and 5(c).

The aforementioned results demonstrate that the approximation of Eq. (3) is more effective when nearing the steady state. If the initial state closely aligns with the steady state, the approximation could represent the entire evolution process of the populations well. Nevertheless, if the initial state significantly deviates from the steady state, the approximation only becomes reliable after a long period.

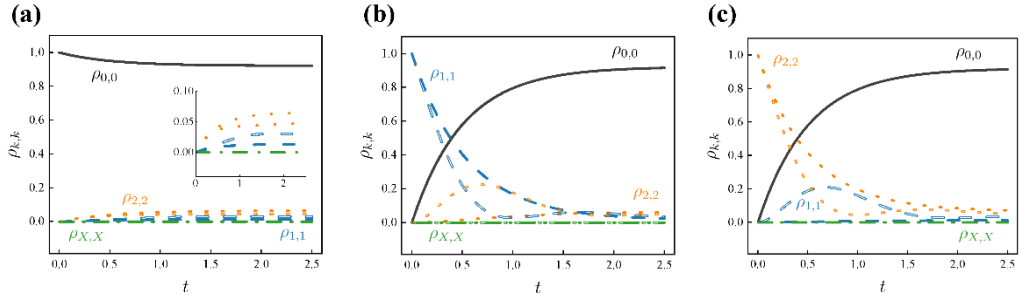

Fig. 6. The theoretical populations evolution of simplified and unsimplified models for different initial states, $\rho_S(0)=|\varepsilon_0\rangle\langle\varepsilon_0|$ (a), $\rho_S(0)=|\varepsilon_1\rangle\langle\varepsilon_1|$ (b), and $\rho_S(0)=|\varepsilon_2\rangle\langle\varepsilon_2|$ (c). The black single (double) solid line, blue single (double) dashed line, orange single (double) dotted line, and green single (double) dash-dotted line illustrates the evolution of $\rho_{0,0}$, $\rho_{1,1}$, $\rho_{2,2}$, and $\rho_{X,X}$ of the simplified (unsimplified) model, respectively. The inset in panel (a) provides a detailed view of the changes in $\rho_{1,1}$, $\rho_{2,2}$, and $\rho_{X,X}$ from $t=0$ to $t=2.5$.

## 4 Conclusions

In this study, we have empirically demonstrated the steady-state behavior of the three-level QHE by designing and executing superconducting quantum circuits. After applying GEM techniques, the results obtained from the quantum circuit model, when executed on a real quantum device, were found to be in close agreement with theoretical predictions. This validates the effectiveness of the proposed circuit model. Furthermore, we also delved into the intricacies of the circuit models, discussing the influence of different initial states and the impact of simplification. Overall, our study presents a streamlined methodology for experimentally exploring the thermodynamics of three-level QHEs, which reduces the complexity of experimental procedures and associated costs.

## Acknowledgement

This work was supported by the Taishan Scholar Project (Grand No. tsqn202103142), Natural Science Foundation of Shandong Province (No. ZR2021QE033).

## Dada Availability

The data that support the findings of this study are available from the corresponding author upon reasonable request.